\documentclass[10pt,journal]{IEEEtran}
\IEEEoverridecommandlockouts
\usepackage{cite}
\usepackage{amsmath,amssymb,amsfonts}
\usepackage{algorithmic}
\usepackage{graphicx}
\usepackage{textcomp}
\usepackage{xcolor}
\usepackage{subfigure}
\usepackage[section]{placeins}
\usepackage{geometry}
\newtheorem{remark}{\text{Remark}}
\newtheorem{lemma}{\text{Lemma}}

\begin{document}

\title{User Power Measurement Based IRS Channel Estimation via Single-Layer Neural Network}

\author{\IEEEauthorblockN{He Sun\IEEEauthorrefmark{1},
                     Weidong Mei\IEEEauthorrefmark{4},
                     Lipeng Zhu\IEEEauthorrefmark{1},
                     and Rui Zhang\IEEEauthorrefmark{2}\IEEEauthorrefmark{3}\IEEEauthorrefmark{1}} \\ 
\IEEEauthorblockA{\IEEEauthorrefmark{1}Department of Electrical and Computer Engineering, National University of Singapore, Singapore 117583.} \\ 
\IEEEauthorblockA{\IEEEauthorrefmark{2}School of Science and Engineering, Shenzhen Research Institute of Big Data.} \\ 
\IEEEauthorblockA{\IEEEauthorrefmark{3}Chinese University of Hong Kong, Shenzhen, Guangdong, China 518172.} \\ 
\IEEEauthorblockA{\IEEEauthorrefmark{4}National Key Laboratory of Science and Technology on Communications, \\
University of Electronic Science and Technology of China, China 611731. \\
E-mail: {sunele@nus.edu.sg, wmei@uestc.edu.cn, zhulp@nus.edu.sg, elezhang@nus.edu.sg}}
\thanks{This paper has been accepted by GLOBECOM 2023.} 
}

\maketitle

\begin{abstract}
One main challenge for implementing intelligent reflecting surface (IRS) aided communications lies in the difficulty to obtain the channel knowledge for the base station (BS)-IRS-user cascaded links, which is needed to design high-performance IRS reflection in practice. Traditional methods for estimating IRS cascaded channels are usually based on the additional pilot signals received at the BS/users, which increase the system training overhead and also may not be compatible with the current communication protocols. To tackle this challenge, we propose in this paper a new single-layer neural network (NN)-enabled IRS channel estimation method based on only the knowledge of users' individual received signal power measurements corresponding to different IRS random training reflections, which are easily accessible in current wireless systems. To evaluate the effectiveness of the proposed channel estimation method, we design the IRS reflection for data transmission based on the estimated cascaded channels in an IRS-aided multiuser communication system. Numerical results show that the proposed IRS channel estimation and reflection design can significantly improve the minimum received signal-to-noise ratio (SNR) among all users, as compared to existing power measurement based designs.
\end{abstract}

\section{Introduction}
Intelligent reflecting surface (IRS) has recently emerged as a candidate technology for the future six-generation (6G) wireless communication systems due to its capability of realizing smart and reconfigurable propagation environment cost-effectively \cite{WQQTutorial}. Specifically, an IRS consists of a large number of passive reflecting elements with independently tunable reflection coefficients, which can be jointly designed to alter the phase and/or amplitude of its incident signal to achieve high performance passive beamforming for various purposes, such as signal boosting, interference cancellation, target sensing, etc\cite{WQQTutorial,surveyapp,JSAC22}.

To this end, IRS passive beamforming or in general passive reflection should be properly designed. In the existing literature, there are two main approaches for IRS passive beamforming design, which are based on channel estimation pilots and user signal power measurements, respectively. In the former approach, the cascaded base station (BS)-IRS-user/user-IRS-BS channels are first estimated based on the downlink/uplink pilots received at the users/BS with time-varying IRS training reflections, and then the IRS reflection for data transmission is optimized based on the estimated IRS cascaded channels\cite{IRSCESurvey1,IRSCESurvey2,mei,DDNN}. Alternatively, the authors in \cite{YWJSAC} proposed to train a deep neural network (NN) to directly learn the mapping from the received pilot signals to the optimal IRS reflection. However, the above pilot-based designs require additional training pilots for IRS channel estimation or NN training, which not only increases the system training overhead but also may not be compatible with the current cellular transmission protocols that cater to the user-BS direct channel (without IRS) estimation only. To efficiently integrate IRS into current wireless systems without the need of changing their protocols, the latter approach designs IRS reflection for data transmission based on the received (pilot or data) signal power measurements at each user's receiver with time-varying IRS reflections, which can be easily obtained in existing wireless systems.
For example, passive beam training for IRS-aided millimeter-wave (mmWave) systems\cite{You1,You2} and conditional sample mean (CSM)-based IRS reflection for IRS-aided sub-6 GHz systems\cite{CSM} have been proposed. In particular, it was shown in \cite{CSM} that in the single-user case, the CSM method can achieve an IRS passive beamforming gain in the order of the number of IRS reflecting elements, which is identical to that under perfect channel state information (CSI)\cite{WQQTutorial}. However, the number of random IRS reflections needed for CSM to obtain sufficient user power measurement samples is very large (hundreds or even thousands) for even the single-user case, which still results in high implementation overhead and large training delay. The fundamental reason for CSM's low efficiency lies in its lack of IRS channel information extraction from the users' power measurements.

\begin{figure}[t]
\begin{center}
\centering
\includegraphics[scale=0.3]{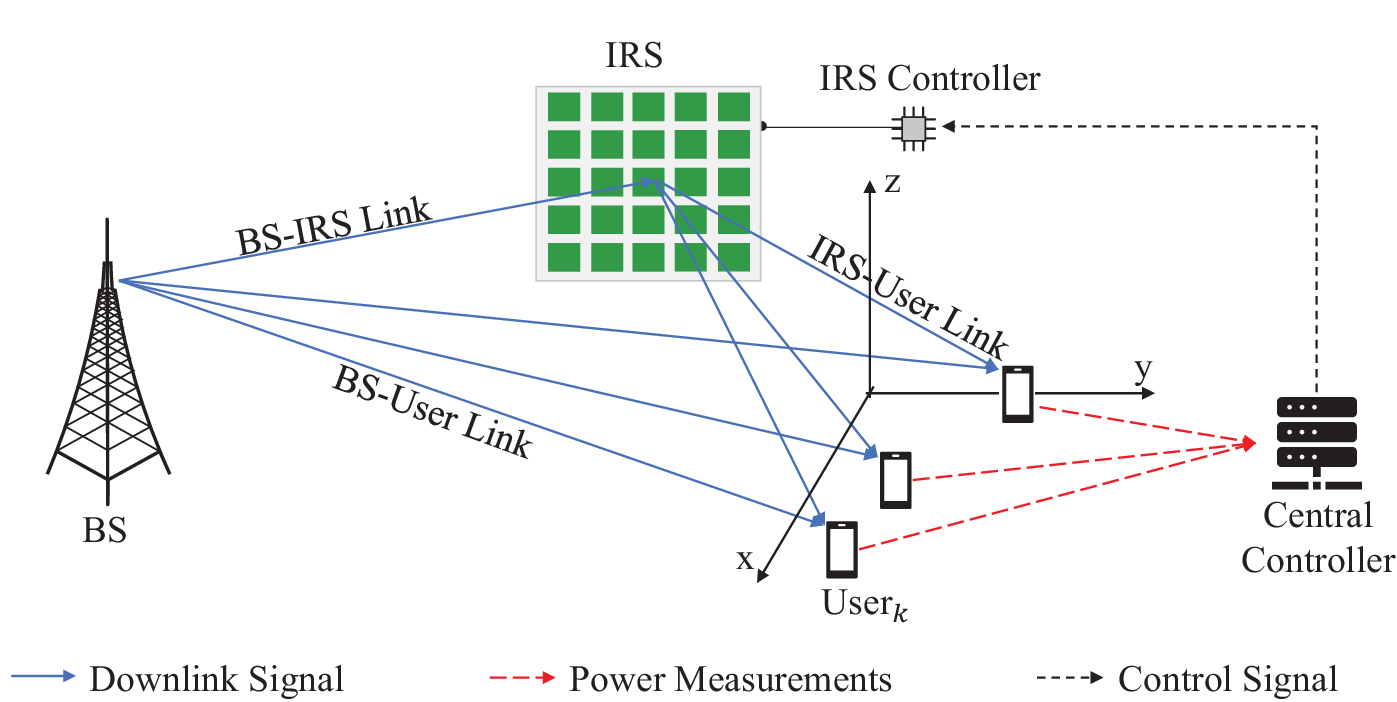} 
\caption{IRS-aided multicasting with users' power measurements.}
\label{fig001}\end{center}
\end{figure}
In this paper, we propose a new IRS cascaded channel estimation and IRS reflection design method based on users' power measurements similar to CSM in \cite{CSM}. However, different from CSM, we first estimate the IRS cascaded channels based on user power measurements and then design the IRS reflection for data transmission based on the estimated channels. This thus overcomes the aforementioned inefficacy of CSM due to the lack of channel information extraction. In particular, our proposed IRS channel estimation method based on user power measurements leverages a simple single-layer NN formulation. Specifically, we first reveal that for any given IRS reflection, the received signal power at each user can be equivalently modeled as the output of a single-layer NN, with its weights corresponding to the coefficients of the cascaded BS-IRS-user channel. Inspired by this, we optimize the weights of the single-layer NN to minimize the mean squared error (MSE) between its output and each user's power measurement via the stochastic gradient descent method, thereby estimating the cascaded BS-IRS-user channel. Next, to evaluate the effectiveness of the proposed channel estimation method, we design the IRS reflection for data transmission based on the estimated cascaded channels in an IRS-aided multiuser multicast communication system, as shown in Fig. \ref{fig001}. We aim to optimize the IRS reflection to maximize the minimum received signal-to-noise ratio (SNR) among all users and solve this problem efficiently by applying various optimization techniques. Numerical results show that the proposed IRS channel estimation and IRS reflection design can yield much better performance than existing user power measurement based schemes such as CSM.

\emph{Notations}: Scalars, vectors and matrices are denoted by lower/upper case, boldface lower case and boldface upper case letters, respectively. For any scalar/vector/matrix, $(\cdot)^*$, $(\cdot)^T$ and $(\cdot)^H$ respectively denote its conjugate, transpose and conjugate transpose. $\mathbb{C}^{n \times m}$ and $\mathbb{R}^{n \times m}$ denote the sets of $n \times m$ complex and real matrices, respectively. $\parallel\cdot\parallel$ denotes the Euclidean norm of a vector, and $\mid\cdot\mid$ denotes the cardinality of a set or the amplitude of a complex scalar. $j=\sqrt{-1}$ denotes the imaginary unit. $\rm Re(\cdot)$ and $\rm Im(\cdot)$ denote the real and imaginary parts of a complex vector/number, respectively. $\boldsymbol{V}\succeq \textbf{0}$ indicates that $\boldsymbol{V}$ is a positive semidefinite matrix. $\text{Tr}(\cdot)$ denotes the trace of a matrix. The distribution of a circularly symmetric complex Gaussian (CSCG) random variable with zero mean and covariance $\sigma^2$ is denoted by $\mathcal{CN}(0,\sigma^2)$.\vspace{-3pt}

\section{System Model and Problem Formulation}

As shown in Fig. \ref{fig001}, we consider an IRS-aided multicast communication system, where a single-antenna BS (or multi-antenna BS with fixed downlink precoding) transmits a common message to $K$ single-antenna users (or independent messages to different users over orthogonal frequency bands), with the help of an IRS consisting of $N$ reflecting elements. It is assumed that there is a central controller in the system (the BS or another dedicated unit) which can collect the users' received signal power measurements and thereby optimize the IRS passive reflection. Let $U_k$ denote the $k$-th user, $k \in {\cal{K}} \triangleq {\{1,2,...,K\}}$. In this paper, we consider quasi-static block-fading channels and focus on a given fading block, during which all the channels involved are assumed to be constant. The baseband equivalent channel from the BS to the IRS, that from the BS to $U_k$, and that from the IRS to $U_k$ are denoted by $\boldsymbol{h}_{BI} \in {\mathbb{C}}^{N \times 1}$, $h_{B{U_k}} \in {\mathbb{{C}}}$ and $\boldsymbol{h}_{I{U_k}}^H \in {\mathbb{C}}^{1 \times N}$, respectively. Let $\mathbf{\Theta}=\text{diag}(e^{j{\theta}_1},...,e^{j{\theta}_N})$ denote the reflection matrix of the IRS, where ${\theta}_i$ denotes the phase shift of its $i$-th reflecting element, $1 \leq i \leq N$. Due to the hardware constraints, we consider that the phase shift of each reflecting element can only take a finite number of discrete values in the set $\Phi_\alpha = \{\omega, {2\omega},{3\omega},...,2^\alpha{\omega}\}$, where $\alpha$ is the number of bits used to uniformly quantize the continuous phase shift in $(0, 2\pi]$, and $\omega = \frac{2\pi}{2^\alpha}$\cite{WdIRS}. Let $P$ denote the transmit power of the BS. The effective channel from the BS to $U_k$ is expressed as\setlength{\baselineskip}{0.95\baselineskip}
\begin{equation}\label{eqs01001}
  g_k = \sqrt{P}\left({h}_{B{U_k}} + {\boldsymbol{h}_{I{U_k}}^H}\mathbf{\Theta}\boldsymbol{h}_{BI}\right), \ k \in {\mathcal{K}},
\end{equation}
where we have incorporated the effect of the BS transmit power $P$ into the BS-$U_k$ effective channel, since it may be practically unknown to the central controller. Let ${\bar{\boldsymbol{v}}}^H=\left[e^{j{\theta}_1},...,e^{j{\theta}_N}\right]$ denote the passive reflection of the IRS, and $\bar{\boldsymbol{h}}_k={\text{diag}}({\boldsymbol{h}_{I{U_k}}^H}){\boldsymbol{h}_{BI}}$ denote the cascaded BS-IRS-$U_k$ channel. As such, the channel in (\ref{eqs01001}) can be simplified as\vspace{-3pt}
\begin{equation}\label{eqs0203}
  g_k = \sqrt{P}{{h}}_{B{U_k}} + \sqrt{P}\bar{\boldsymbol{v}}^H\bar{\boldsymbol{h}}_k, \ k \in {\mathcal{K}}.
\end{equation}
By extending the IRS passive reflection vector into $\boldsymbol{v}^H=\left[1, \bar{\boldsymbol{v}}^H\right]$ and stacking the direct and cascaded BS-$U_k$ channels into $\boldsymbol{h}_k^H=\sqrt{P}\left[{{h}_{B{U_k}}^*}, \bar{\boldsymbol{h}}_k^H\right]$, the baseband equivalent channel in (\ref{eqs0203}) can be further simplified as
\begin{equation}\label{eqs001}
  g_k = \boldsymbol{v}^H{\boldsymbol{h}_k}, \ k \in {\mathcal{K}}.
\end{equation}
Let $s \in \mathbb{C}$ denote the transmitted symbol (pilot or data) at the BS with $\left\vert s \right\vert^2=1$. Hence, the received signal at $U_k$ is given by
\begin{equation}\label{eqs004}
  y_{k} = {g_{k}}{s} + \emph{n}_k, \ k \in {\mathcal{K}},
\end{equation}
where $n_k\sim\mathcal{CN}(0,\sigma^2)$ denotes the complex additive white Gaussian noise (AWGN) at $U_k$ with power ${\sigma}^2$. Accordingly, the received SNR at $U_k$ is
\begin{equation}\label{eqs005}
  \text{SNR}_k = \frac{{\left\vert g_{k}\right\vert}^2}{{\sigma}^2} = \frac{{\boldsymbol{v}}^H{\boldsymbol{G}_k}{\boldsymbol{v}}}{{\sigma}^2}, \ k \in {\mathcal{K}},
\end{equation}
where $\boldsymbol{G}_k= {\boldsymbol{h}_k}\boldsymbol{h}_k^H$ denotes the covariance matrix of $\boldsymbol{h}_k$. In this paper, we aim to optimize the IRS passive reflection to maximize the minimum received SNR among all $K$ users. The associated optimization problem is thus formulated as
\begin{subequations}\label{eqs004}
\begin{align}
  \text{(P1):} &\max_{\boldsymbol{v}} \ \min_{{{k \in \mathcal{K}}}} \ {\boldsymbol{v}}^H{\boldsymbol{G}_k}{\boldsymbol{v}}\label{Za}\\
& \text{s.t.} \ \ \ {\theta}_i\in \Phi_\alpha, i=1,...,N, \label{Zb}
\vspace{-3pt}\end{align}
\end{subequations}
where the constant scalar $1/\sigma^2$ is omitted in (\ref{Za}). Problem (P1) is a discrete optimization problem and can be solved by a variety of methods (see e.g., \cite{WdIRS,OptYG}), given that the perfect channel knowledge (i.e., $\boldsymbol{h}_k$ or $\boldsymbol{G}_k$, $k \in {\mathcal{K}}$) is available. However, perfect CSI is difficult to acquire in practice for IRS cascaded channels \cite{WQQTutorial}. Thus, in this paper, we propose an efficient single-layer NN-enabled method to estimate $\boldsymbol{h}_k$, $k \in {\mathcal{K}}$, based on the users' individual power measurements and then optimize $\boldsymbol{v}$ via solving (P1), as detailed in the next section.\vspace{-1.9pt}

\section{Single-Layer NN-enabled IRS Channel Estimation and IRS Reflection Optimization}

\subsection{User Power Measurement}
In current wireless systems (e.g., cellular and WiFi), user terminals are able to measure the average power of their received signals, referred to as the reference signal received power (RSRP). By leveraging this capability, we propose to estimate the IRS cascaded channels based on users' power measurements (to be presented in the next subsection). Specifically, we assume that the IRS applies randomly generated phase shifts of its reflecting elements (subject to (\ref{Zb})) to reflect the BS's signals to all $K$ users simultaneously. Let $M$ and ${\boldsymbol{v}}_m$, $m \in {\cal{M}}\triangleq \left\{1,2,...,M\right\}$, denote the number of random reflection sets generated and its $m$-th reflection set, respectively. In the meanwhile, the users independently measure the power of their received signals corresponding to each IRS reflection set and send the results to the central controller (see Fig.\ref{fig001}). For each reflection set, we consider that $U_k$ takes $Q$ samples of its received signal to calculate the RSRP based on them. Note that in practice, it usually holds that $Q \gg 1$ since the IRS's reflection switching rate is usually much lower than the symbol rate of each user. Thus, the power measurement of $U_k$ under the IRS's $m$-th reflection set is given by
\begin{equation}\label{eqs}
{\bar p}_k(\boldsymbol{v}_m) = \frac{1}{Q}\sum_{q=1}^Q{\left|{g_{k,m}}{s} + \emph{n}_k(q)\right|^2}, \ k \in {\mathcal{K}}, \ m \in {\cal{M}},
\vspace{-3pt}\end{equation}
where $g_{k,m}=\boldsymbol{v}_m^H\boldsymbol{h}_k$ and $\emph{n}_k(q)\sim\mathcal{CN}(0,\sigma^2)$ denotes the $q$-th sampled AWGN at $U_k$. Let $\mathcal{P}_k = \left[{\bar p}_k(\boldsymbol{v}_1),{\bar p}_k(\boldsymbol{v}_2),...,{\bar p}_k(\boldsymbol{v}_M) \right]$ denote the collection of $U_k$'s received signal power measurements under the $M$ reflection sets of the IRS. After the above power measurements, each user $U_k$ reports $\mathcal{P}_k$ to the central controller, which estimates ${\boldsymbol{h}}_k, k \in {\cal{K}}$ as presented next.\vspace{-2.9pt}
\subsection{NN-enabled Channel Estimation}\label{secIIIA}
For any given IRS reflection set $\boldsymbol{v}$, the desired signal power at each user $U_k$ is given by
\begin{equation}\label{eqs03007}
p_k(\boldsymbol{v})={\left|{\boldsymbol{v}^H{\boldsymbol{h}_k}} \right|}^2.
\end{equation}
It is worth mentioning that if the number of samples $Q$ is sufficiently large, we have ${\bar p}_k(\boldsymbol{v}) \approx {p}_k(\boldsymbol{v}) + \sigma^2$. Thus, we aim to estimate $\boldsymbol{h}_k, k \in {\cal{K}}$ based on ${\bar p}_k(\boldsymbol{v}_m), m \in {\cal{M}}$. To this end, note that (\ref{eqs03007}) can be modeled as a single-layer NN, explained as follows. In particular, this NN takes the reflection pattern $\boldsymbol{v}$ and the cascaded channel ${\boldsymbol{h}_k}$ as its input and weights, respectively, while the nonlinear activation function at the output layer is the squared amplitude of ${\boldsymbol{v}^H{\boldsymbol{h}_k}}$, as given in (\ref{eqs03007}). However, as both $\boldsymbol{v}$ and ${\boldsymbol{h}_k}$ are complex numbers in general, such a single-layer NN requires the implementation in the complex domain. To avoid this issue, we express (\ref{eqs03007}) equivalently in the real domain as
\begin{equation}\label{eqs03008}
p_k(\boldsymbol{v})={\left\vert {\boldsymbol{v}^H{\boldsymbol{h}_k}} \right\vert}^2 = \left\|{\boldsymbol{x}^T \boldsymbol{R}_k}\right\|^2,
\end{equation}
where $\boldsymbol{x}$ consists of the real and imaginary parts of $\boldsymbol{v}$, i.e., $\boldsymbol{x}^T = \left[{\begin{array}{*{20}{c}}
{{\mathop{\rm Re}\nolimits} \left( {{\boldsymbol{v}^T}} \right)},&{{\mathop{\rm Im}\nolimits} \left( {{\boldsymbol{v}^T}} \right)}
\end{array}}\right]$, and $\boldsymbol{R}_k$ denotes the real-valued cascaded channel, i.e.,
\begin{equation}\label{eqs0300801}
\boldsymbol{R}_k=\left[ {\begin{array}{*{20}{c}}
{ {{\mathop{\rm Re}\nolimits} \left( \boldsymbol{h}_k \right)} }&{{{{\mathop{\rm Im}\nolimits} \left( \boldsymbol{h}_k \right)}}}\\
{ {{{\mathop{\rm Im}\nolimits} \left( \boldsymbol{h}_k \right)}}}&{{-{{\mathop{\rm Re}\nolimits} \left( \boldsymbol{h}_k \right)}}}
\end{array}} \right]\in \mathbb{R}^{{(2N+2)} \times 2}.
\end{equation}

\begin{figure}[t]
\begin{center}
\centering
\includegraphics[scale=0.3]{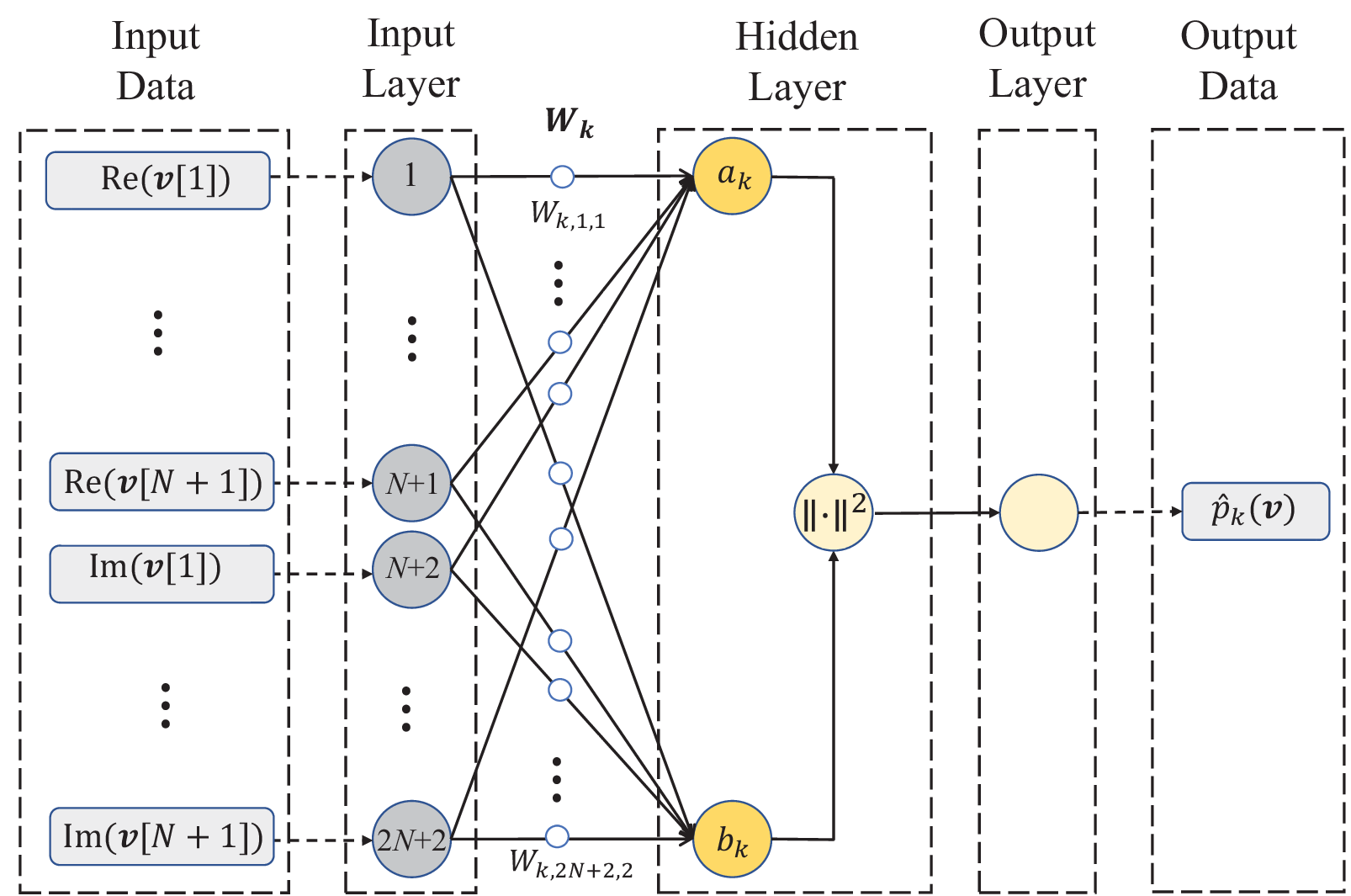} 
\caption{Single-layer NN architecture for $U_k$. }
\vspace{-16pt}
\label{fig002}\end{center}
\end{figure}

Based on (\ref{eqs03008}), we can construct an equivalent single-layer NN to (\ref{eqs03007}) in the real-number domain. Specifically, as shown in Fig. \ref{fig002}, the input of this single-layer NN is $\boldsymbol{x}$. Let $W_{k,i,j}$ denote the weight of the edge from the $i$-th input to the $j$-th neuron in the hidden layer, with $i=1,2,...,2N+2$ and $j=1,2$. The two neurons at the hidden layer of this NN are given by
\begin{equation}\label{eqs03009}
  \left[ {\begin{array}{*{20}{c}}
{a_k}\\
{b_k}
\end{array}} \right]^T = \boldsymbol{x}^T \boldsymbol{W}_k,
\end{equation}
where $\boldsymbol{W}_k\in \mathbb{R}^{(2N+2) \times 2}$ denotes the weight matrix of this NN, with $W_{k,i,j}$ being its entry in the $i$-th row and the $j$-th column.

Finally, the activation function at the output layer is given by the squared norm of (\ref{eqs03009}), and the output of this NN is
\begin{equation}\label{eqs03011}
  {{\hat p}_k}(\boldsymbol{v}) = a_k^2 + b_k^2 = \left\|{\boldsymbol{x}^T \boldsymbol{W}_k}\right\|^2.
\end{equation}
By comparing (\ref{eqs03011}) with (\ref{eqs03008}), it is noted that this real-valued NN can imitate the received signal power by $U_k$. In particular, if ${\boldsymbol{R}}_k={\boldsymbol{W}}_k$, we have $\hat p_k(\boldsymbol{v})=p_k(\boldsymbol{v})$. Motivated by this, we propose to recover ${\boldsymbol{R}}_k$ (and ${\boldsymbol{h}}_k$) by estimating the weight matrix ${\boldsymbol{W}}_k$ via training this single-layer NN. To this end, we consider that ${\boldsymbol{W}}_k$ takes a similar form to ${\boldsymbol{R}}_k$ in (\ref{eqs0300801}), i.e.,
\begin{equation}\label{eqs03010}
  {\boldsymbol{W}}_k=\left[ {\begin{array}{*{20}{c}}
{ {{\boldsymbol{w}_{1,k}}} }&{{{\boldsymbol{w}}_{2,k}}}\\
{ {{\boldsymbol{w}_{2,k}}}}&{{-{\boldsymbol{w}_{1,k}}}}
\end{array}} \right],
\end{equation}
where ${\boldsymbol{w}}_{1,k} \in \mathbb{R}^{N+1}$ and ${\boldsymbol{w}}_{2,k}\in \mathbb{R}^{N+1}$ correspond to ${{{\mathop{\rm Re}\nolimits} \left( \boldsymbol{h}_k \right)} }$ and ${{{\mathop{\rm Im}\nolimits} \left( \boldsymbol{h}_k \right)} }$ in (\ref{eqs0300801}), respectively.

With (\ref{eqs03010}), we present the following lemma.\setlength{\baselineskip}{0.99\baselineskip}
\begin{lemma}If
\begin{equation}\label{eqs03018}
  \|\boldsymbol{x}^T {\boldsymbol{W}}_k\|^2 = \|{\boldsymbol{x}^T \boldsymbol{R}_k}\|^2
\end{equation}
holds for any $\boldsymbol{x} \in \mathbb{R}^{2N+2}$, we have ${\boldsymbol{h}}_k={\boldsymbol{w}}_k e^{j\phi_k}$, where ${\boldsymbol{w}}_k={{\boldsymbol{w}}_{1,k}} + j {\boldsymbol{w}}_{2,k}$, $\phi_k \in [0,2\pi)$ denotes an arbitrary phase.
\end{lemma}

\begin{IEEEproof}
By substituting (\ref{eqs03010}) into the left-hand side of (\ref{eqs03018}), we have
\begin{equation}\label{eqs0301801}
  \|\boldsymbol{x}^T {\boldsymbol{W}}_k\|^2 = {\left| {\boldsymbol{v}^H{\boldsymbol{w}}_k} \right|}^2={\boldsymbol{v}^H{\boldsymbol{w}}_k}{{\boldsymbol{w}}_k^H\boldsymbol{v}}.
\end{equation}
Next, by substituting (\ref{eqs03008}) and (\ref{eqs0301801}) into (\ref{eqs03018}), we have
\begin{equation}\label{eqs03019}
{\boldsymbol{v}^H{\boldsymbol{w}}_k}{{\boldsymbol{w}}_k^H\boldsymbol{v}}={\left| {\boldsymbol{v}^H{\boldsymbol{h}_k}} \right|}^2={{\boldsymbol{v}}^H{{{{\boldsymbol{h}}}}_k{{{\boldsymbol{h}}}}_k^H}{\boldsymbol{v}}}, \ \forall \boldsymbol{v} \in {\mathbb{C}}^{N+1},
\end{equation}
which implies that
\begin{equation}\label{eqs03021}
{\boldsymbol{v}}^H\left({{{\boldsymbol{h}}}}_k{{{\boldsymbol{h}}}}_k^H - {\boldsymbol{w}}_k{\boldsymbol{w}}_k^H\right){\boldsymbol{v}} = 0, \ \forall {\boldsymbol{v}} \in {\mathbb{C}}^{N+1}.
\end{equation}
For (\ref{eqs03021}) to hold for any ${\boldsymbol{v}} \in {\mathbb{C}}^{N+1}$, it should be satisfied that ${{{\boldsymbol{h}}}}_k{{{\boldsymbol{h}}}}_k^H = {\boldsymbol{w}}_k{\boldsymbol{w}}_k^H$. As such, we have ${\boldsymbol{h}}_k={\boldsymbol{w}}_k e^{j\phi_k}$. The proof is thus completed.
\end{IEEEproof}

It follows from Lemma 1 that we can estimate ${\boldsymbol{h}}_k$ by training the single-layer NN in Fig. \ref{fig002} to estimate ${\boldsymbol{W}}_k$ first. Although we cannot derive the exact ${\boldsymbol{h}}_k$ due to the presence of the unknown phase $\phi_k$, the objective function of (P1) only depends on the channel covariance matrix ${\boldsymbol{G}}_k$, and we have ${\boldsymbol{G}}_k={{{\boldsymbol{h}}}}_k{{{\boldsymbol{h}}}}_k^H = {\boldsymbol{w}}_k{\boldsymbol{w}}_k^H, k \in {\cal{K}}$. As such, the unknown common phase does not affect the objective function of (P1). It should also be mentioned that Lemma 1 requires that (\ref{eqs03018}) holds for any ${\boldsymbol{x}} \in {\mathbb{R}}^{2N+2}$ or ${\boldsymbol{v}} \in {\mathbb{C}}^{N+1}$. However, due to (\ref{Zb}), the discrete IRS passive reflection set can only take a finite number of values in a subspace of ${\mathbb{C}}^{N+1}$, and ${\boldsymbol{h}}_k={\boldsymbol{w}}_k e^{j\phi_k}$ may not always hold in such a subspace. Nonetheless, the proposed design is still effective, as will be explained in Remark 1 later.

To estimate ${{{\boldsymbol{w}}}}_k$ or ${{{\boldsymbol{W}}}}_k$, we can train the NN in Fig. \ref{fig002} by using the stochastic gradient descent method to minimize the MSE between its output and the training data. In particular, we can make full use of each user's power measurements, i.e., ${\cal P}_k, k \in {\cal K}$, as the training data. Specifically, we divide them into two data sets, namely, the training set and validation set. The training set consists of $M_0$ $(M_0<M)$ entries of ${\cal P}_k$, while the remaining $M-M_0$ entries of ${\cal P}_k$ are used as the validation set to evaluate the model fitting accuracy.
Accordingly, the MSE for the training data is set as the following loss function,
\begin{equation}\label{eqs0310}
  {\cal L}_{\boldsymbol{W}_k} = \frac{1}{M_0} \sum_{m=1}^{M_0} {\left({\hat {{{p}}}}_k(\boldsymbol{v}_m)-{{\bar{p}}_k}(\boldsymbol{v}_m)\right)}^2.
\end{equation}
Given this loss function, we can use the backward propagation \cite{rumelhart1986learning} to iteratively update the NN weights. Specifically, with (\ref{eqs03010}), the weight matrix ${\boldsymbol{W}_k}$ can be expressed by a vector $\boldsymbol{\gamma}_k =\left[{{\boldsymbol{w}}_{1,k}}^T, {\boldsymbol{w}}_{2,k}^T\right]^T \in \mathbb{R}^{2N+2}$. Let $\boldsymbol{\gamma}_{k,t}$ denote the updated value of $\boldsymbol{\gamma}_k$ after the $t$-th iteration. As such, the iteration proceeds as
\begin{equation}\label{eqs0312}
  {\boldsymbol{\gamma}_{k,t+1}} = {\boldsymbol{\gamma}_{k,t}} - \rho F\left({\boldsymbol{\gamma}_{k,t}}\right),
\end{equation}
where $\rho>0$ denotes the learning rate, and $F\left({\boldsymbol{\gamma}_{k}}\right)=\frac{\partial{{\cal L}_{{\boldsymbol{W}}_k}}}{\partial{{{\boldsymbol{\gamma}}_k}}}$ denotes the derivative of the loss function ${\cal L}_{\boldsymbol{W}_k}$ with respect to $\boldsymbol{\gamma}_k$.
Here, $F\left({\boldsymbol{\gamma}_{k}}\right)$ can be calculated using the chain rule,
\begin{equation}\label{eqs000001}
  F\left(\boldsymbol{\gamma}_k\right) = \frac{\partial{{\cal L}_{{\boldsymbol{W}}_k}}}{\partial{{{\hat{p}}_k}}}\left[\frac{\partial{{{\hat{p}}_k}}}{\partial{{{{a}}_k}}}, \ \frac{\partial{{{\hat{p}}_k}}}{\partial{{{{b}}_k}}}\right]
  \left[\frac{\partial{{{{a}}_k}}}{\partial{{{{\boldsymbol{\gamma}}}_k}}}, \ \frac{\partial{{{{b}}_k}}}{\partial{{{{\boldsymbol{\gamma}}}_k}}}\right]^T,
\end{equation}
where $\frac{\partial{{\cal L}_{{\boldsymbol{W}}_k}}}{\partial{{{\hat{p}}_k}}}$ can be calculated based on (\ref{eqs0310}), while the other four derivatives in (\ref{eqs000001}) can be calculated based on (\ref{eqs03009}) as
\begin{align}
\frac{\partial{{{{a}}_k}}}{\partial{{{{\boldsymbol{\gamma}}}_k}}} & = \left[ {1,
{\cos \left( {{\theta _1}} \right)},\!\cdots\!, {\cos \left( {{\theta _{N}}} \right)},
{0, -\sin \left( {{\theta _1}} \right)},\!\cdots\!,{-\sin \left( {{\theta _{{N}}}} \right)}} \right]^T,\nonumber\\
\frac{\partial{{{{b}}_k}}}{\partial{{{{\boldsymbol{\gamma}}}_k}}} & = \left[
{0, \sin \left( {{\theta_1}} \right)},\! \cdots\!,{\sin \left( {{\theta_{N}}} \right)},
{1, \cos \left( {{\theta_1}} \right)},\! \cdots\!,{\cos \left( {{\theta_{N}}} \right)} \right]^T, \nonumber\\
\frac{\partial{{{\hat{p}}_k}}}{\partial{{{{a}}_k}}} & = 2{{{a}}_k}, \ \text{and} \ \
\frac{\partial{{{\hat{p}}_k}}}{\partial{{{{b}}_k}}} = 2{{b}}_k.
\end{align}
The NN training process terminates after $Z$ rounds of iterations, and the weight matrix of the NN is determined as
\begin{equation}\label{eqs0311}
  \boldsymbol{W}^{\divideontimes}_k = \arg \min_{1 \le t \le Z} \left(\sum_{m=M_0+1}^{M} {(\hat {p}_{k,t}}(\boldsymbol{v}_m)-{{\bar{p}}_k}(\boldsymbol{v}_m))^2\right),
\end{equation}
based on the validation set, where ${\hat{p}}_{k,t}(\boldsymbol{v}_m)=\|\boldsymbol{x}_m^T {\boldsymbol{W}}_{k,t}\|^2$ denotes the output of the NN after the $t$-th iteration, and $\boldsymbol{W}_{k,t}$ denotes the updated version of $\boldsymbol{W}_k$ after the $t$-th iteration. Based on the above, the complex-valued cascaded channel can be estimated as ${\boldsymbol{w}}^{\divideontimes}_k={{\boldsymbol{w}}^{\divideontimes}_{1,k}} + j {\boldsymbol{w}}^{\divideontimes}_{2,k}$.

\begin{remark}\label{remark1}
In the case with one-bit IRS phase shifts, i.e., $\alpha =1$, the cascaded channel ${\boldsymbol{h}}_k$ may not be estimated as ${\boldsymbol{w}}_k e^{j\phi_k}$. This is because in this case, we have ${\boldsymbol{v}^*}={\boldsymbol{v}}$, which results in
\begin{equation}\label{eqs03023}
  {\boldsymbol{v}^H}{{{\boldsymbol{h}}}}_k{{{\boldsymbol{h}}}}_k^H\boldsymbol{v}={\boldsymbol{v}^H}{\boldsymbol{h}}_k^{*}{\boldsymbol{h}}_k^{T}\boldsymbol{v}.
\end{equation}
Based on (\ref{eqs03021}), we may estimate ${\boldsymbol{h}}_k^*$ as ${\boldsymbol{w}}_k e^{j\phi_k}$, while the actual channel ${\boldsymbol{h}}_k$ should be estimated as ${\boldsymbol{w}}_k^* e^{-j\phi_k}$. However, this does not affect the efficacy of the proposed design, since both estimations lead to the same received signal power due to (\ref{eqs03023}).
\end{remark}
\subsection{IRS Reflection Optimization}\label{SECIIIC}
After estimating ${\boldsymbol{h}}_k, k \in {\cal{K}}$, we can substitute them into (\ref{Za}) and solve (P1) accordingly. Next, we present the optimal and suboptimal algorithms to solve (P1) in the cases of $K=1$ and $K>1$, respectively. First, if $K=1$, it has been shown in \cite{OptAPX} that (P1) can be optimally solved by applying a geometry-based method, and the details are thus omitted.

However, if $K>1$, problem (P1) is generally difficult to be optimally solved. Next, we consider combining the SDR technique\cite{SDR} and the successive refinement method\cite{WdIRS} to solve it. First, let ${\boldsymbol{V}}=\boldsymbol{v}\boldsymbol{v}^H$ denote the covariance matrix of $\boldsymbol{v}$, with $\boldsymbol{V}\succeq \textbf{0}$. Problem (P1) can be equivalently reformulated as
\begin{subequations}\label{eqs03016}
\begin{align}
  \text{(P2):} &\max_{{\boldsymbol{V}}} \ \xi\label{O03016}\\
& \text{s.t.} \ \ \ \text{Tr}({\hat{{\boldsymbol{G}}}_k}{\boldsymbol{V}}) \geq \xi, \ \forall \ k, \label{C0301601}\\
 & \ \ \ \ \ \ \text{rank}({\boldsymbol{V}})=1,\label{O030165}\\
 & \ \ \ \ \ \ {\boldsymbol{V}}\succeq \textbf{0},\label{C0301603}\\
 & \ \ \ \ \ \ \ {\theta}_i\in \Phi_\alpha, \ i=1,...,N\label{C0301604},
\end{align}
\end{subequations}
where $\xi$ is an auxiliary variable. Problem (P2) is still difficult to solve due to the rank-one and discrete-phase constraints in (\ref{O030165}) and (\ref{C0301604}), respectively. Next, we relax both constraints and thereby transform (P2) into a semidefinite programming (SDP) problem, which can be optimally solved by the interior-point algorithm\cite{cvx}. However, the obtained solution may not be rank-one, and its entries may not satisfy the discrete constraint (\ref{C0301604}). In this case, we can apply the Gaussian randomization method jointly with the solution quantization to construct a rank-one solution that satisfies (\ref{C0301604}), denoted as $\hat {\boldsymbol{v}}$.
Based on the initial passive reflection $\hat {\boldsymbol{v}}$, we successively refine $\theta_i$ by enumerating the elements in $\Phi_{\alpha}$, with $\theta_j, j \ne i, j=1,2,...,N$ being fixed, until the convergence is reached.

\subsection{Complexity Analysis}
In the proposed IRS channel estimation and IRS reflection design method, the computational complexity is mainly due to the NN training procedures for channel estimation and the passive reflection optimization for solving (P2). In particular, the training complexity depends on the size of the NN structure. In the NN for $U_k, k \in {\cal K}$, as shown in Fig. \ref{fig002}, the number of neurons is $2$, and the number of weights is $2N+2$, which entails the complexity for all $K$ users in the order of ${\cal O}\left(KN\right)$\cite{haykinneural}. Furthermore, in the passive reflection optimization, the SDR-based initialization incurs the complexity of ${\cal O}\left( (K+N)^{3.5}\right)$, while the successive refinement incurs the complexity of ${\cal O}\left(KN\right)$. Thus, the overall complexity of the proposed design is dominated by the SDR, i.e., ${\cal O}\left( (K+N)^{3.5} \right)$. In practice, we can apply the successive refinement only for solving (P2) with affordable performance loss (to be shown in Section \ref{SECIV} via simulation), while the complexity is decreased significantly to ${\cal O}\left( KN \right)$, thus reducing the overall complexity to linear over $N$.

\section{Numerical Results}\label{SECIV}
\subsection{Simulation Setup}
Consider a three-dimensional Cartesian coordinate system in meter (m) with $K$ users, where the BS is deployed at $(50, -200, 20)$, while the locations of all users are randomly generated in a square area with the coordinates of its four corner points given by $(0,0,0), (10,0,0), (10,10,0)$ and $(0,10,0)$, respectively. The IRS is equipped with a uniform planar array (UPA) and assumed to be parallel to the $y$-$z$ plane, with $N=N_y \times N_z$ reflecting elements, where $N_y$ and $N_z$ denote the numbers of reflecting elements along the axes $y$ and $z$, respectively. We set $N_y=N_z=8$ and half-wavelength spacing for the adjacent IRS reflecting elements. The location of the reference point of the IRS is set as $(-2, -1, 0)$. Let $\beta_{0,k}$, $\beta_{1}$ and $\beta_{2,k}$ denote the path loss (in dB) of the BS-$U_k$, BS-IRS and IRS-$U_k$ channels, respectively, which are set to $\beta_{0,k} = 33 + 37\text{log}_{10}(d_{0,k})$, $\beta_{1}= 30 + 20\text{log}_{10}(d_{1})$ and $\beta_{2,k} = 30 + 20\text{log}_{10}(d_{2,k})$, respectively, with $d_{0,k}$, $d_{1}$ and $d_{2,k}$ denoting the distance from the BS to $U_k$, that from the BS to the IRS, and that from the IRS to $U_k$. We assume Rayleigh fading for the BS-$U_k$ channel, i.e., ${{\emph{{h}}}_{B{U_k}}} = 10^{-\beta_{0,k}/20}\zeta_{k}$,
where $\zeta_{k}$ denotes the small-scale fading following $\mathcal{CN}(0,1)$.
In addition, a multipath channel model is assumed for the BS-IRS and IRS-$U_k$ channels, and the BS-IRS channel is expressed as
\begin{equation}\label{eqs401}
{{{\boldsymbol{h}}}_{BI}} = \sqrt{\frac{\varepsilon_{BI}}{1+\varepsilon_{BI}}}{\boldsymbol{h}}_{LoS} + \sqrt{\frac{1}{1+\varepsilon_{BI}}}{\boldsymbol{h}}_{NLoS},
\end{equation}
where $\varepsilon_{BI}$ is the ratio of the line-of-sight (LoS) path power to that of the non-LoS (NLoS) path. ${\boldsymbol{h}}_{LoS}$ and ${\boldsymbol{h}}_{NLoS}$ denote the LoS and NLoS components, which are respectively given by
\begin{subequations}\label{eqs402}
\begin{align}
  {\boldsymbol{h}}_{LoS} \ & =10^{-\beta_{1}/20}{e^{\frac{-j2\pi {d_{1}}}{\lambda}}}{\boldsymbol{u}_N(\vartheta_{0},\varphi_{0})},\label{Zg}\\
  \boldsymbol{h}_{NLoS} & =\sqrt{\frac{1}{L}}\sum_{l=1}^L{{\kappa_l}\boldsymbol{u}_N(\vartheta_{l},\varphi_{l})},\label{Zh}
\end{align}
\end{subequations}
where $\lambda$ denotes the wavelength. In (\ref{eqs402}), $L$ denotes the number of NLoS multipath components, $\kappa_l$ denotes the amplitude of the $l$-th multipath component following ${\cal{CN}}(0,10^{-\beta_{1}/10})$, and $\boldsymbol{u}_N(\vartheta_{l},\varphi_{l})$ denotes the steering vector of the $l$-th path from the BS to the IRS with $\vartheta_{l} \in [0,\pi]$ and $\varphi_{l}\in [0,\pi]$ denoting the azimuth and the elevation angles of arrival at the IRS in this path, respectively. In particular, let $\boldsymbol{e}(\gamma,n)=[1, e^{-j\pi\gamma}, e^{-j2\pi\gamma},...,e^{-j{(n-1)}\pi\gamma}]^T$ denote the steering vector function of a uniform linear array with $n$ elements and directional cosine $\gamma$. As such, we have ${\boldsymbol{u}_N(\vartheta_{l},\varphi_{l})}  = \boldsymbol{e}({\rm sin}(\vartheta_{l}){\rm sin}(\varphi_{l}), N_y)\otimes\boldsymbol{e}({\rm cos}(\vartheta_{l}), N_z)$, where $\otimes$ denotes the Kronecker product. The IRS-$U_k$ channel can be expressed similarly and we denote by $\varepsilon_{IU_k}$ the ratio of its LoS path power to that of the NLoS counterpart. We set $L=5$, $\varepsilon_{BI}=10$ and $\varepsilon_{IU_k}=1$. The number of power measurements obtained by one user under each IRS reflection set is $Q=10$. The transmit power is $P=30$ dBm, and the noise power is $\sigma^2=-90$ dBm. All results are averaged over $10^3$ realizations of channels and user locations.

\subsection{Benchmark Schemes}
We adopt the CSM\cite{CSM} and random-max sampling (RMS)\cite{tao2020intelligent} methods as benchmark schemes, both of which design the IRS passive reflection for data transmission based on the users' power measurements, but without estimating the IRS cascaded channels by further exploiting the power measurements. Specifically, the RMS method sets the IRS reflection as the one that maximizes the minimum received signal power among all users over $M$ random IRS reflection sets, i.e.,
\begin{equation}\label{s3001}
  {\boldsymbol{v}}^{\text{RMS}} = \boldsymbol{v}_{m^{\star}}, \ \ \ \text{with} \ \ \ {m^{\star}}=\arg{{\max_{m \in \mathcal{M}} \ {\min_{k \in \mathcal{K}}{{\bar{p}}_k(\boldsymbol{v}_m)}}}}.
\end{equation}

Moreover, the CSM method first calculates the sample mean of the minimum power measurement among all users conditioned on $\theta_i=\psi, \psi \in \Phi_{\alpha}$, i.e.,
\begin{equation}\label{s3002}
  {{\mathbb{E}}}[p|\theta_i=\psi] = \frac{1}{\left| \mathcal{A}_{i}(\psi) \right|} \sum_{\boldsymbol{v} \in \mathcal{A}_{i}(\psi)}{\min_{k \in \mathcal{K}}{{\bar{p}}_{k}(\boldsymbol{v})}},
\end{equation}
where $\mathcal{A}_{i}(\psi)$ denotes a subset of the $M$ random reflection sets with $\theta_i=\psi$, $i=1,2,...,N$. Finally, the phase shift of the $i$-th reflecting element is set as
\begin{equation}\label{s3003}
  \theta_i^{\text{CSM}} = \arg{\max_{\psi \in \Phi_\alpha}{{{\mathbb{E}}}[p|\theta_i=\psi]}}, \ i=1,...,N.
\end{equation}
In addition, the IRS passive reflection design based on perfect CSI by solving (P2) with ${\hat{\boldsymbol{G}}}_k$ replaced by ${{\boldsymbol{G}}}_k, k \in {\cal{K}}$, is included as the performance upper bound to evaluate the efficacy of the proposed scheme.
\begin{figure}[t]
\begin{center}
\centering
\includegraphics[scale=0.5]{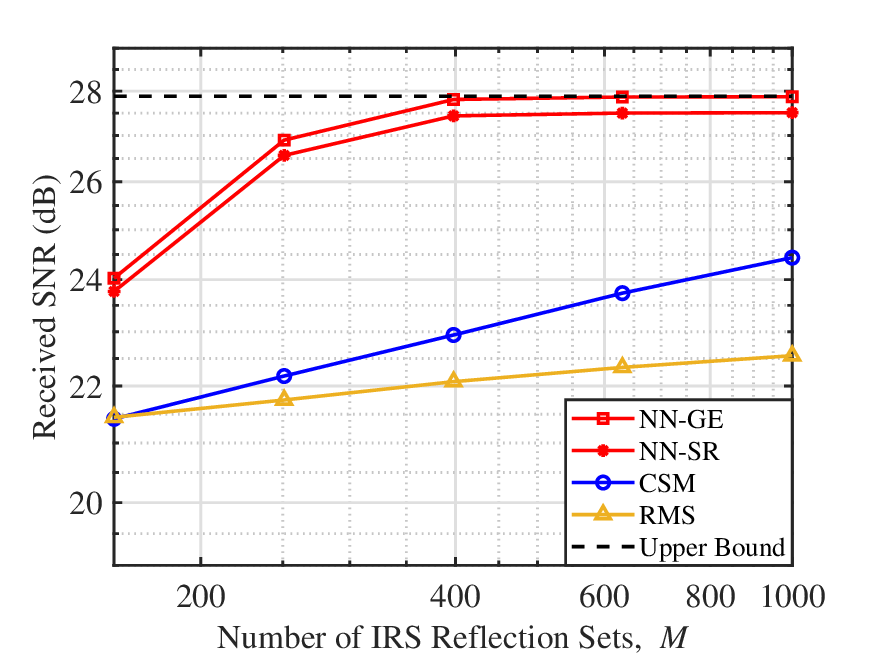} 
\caption{Received SNR versus the number of IRS reflection sets with $K=1$ and $\alpha=1$.}
\label{fig1}\end{center}
\vspace{-10pt}
\end{figure}

\begin{figure}[t]
\begin{center}
\centering
\includegraphics[scale=0.5]{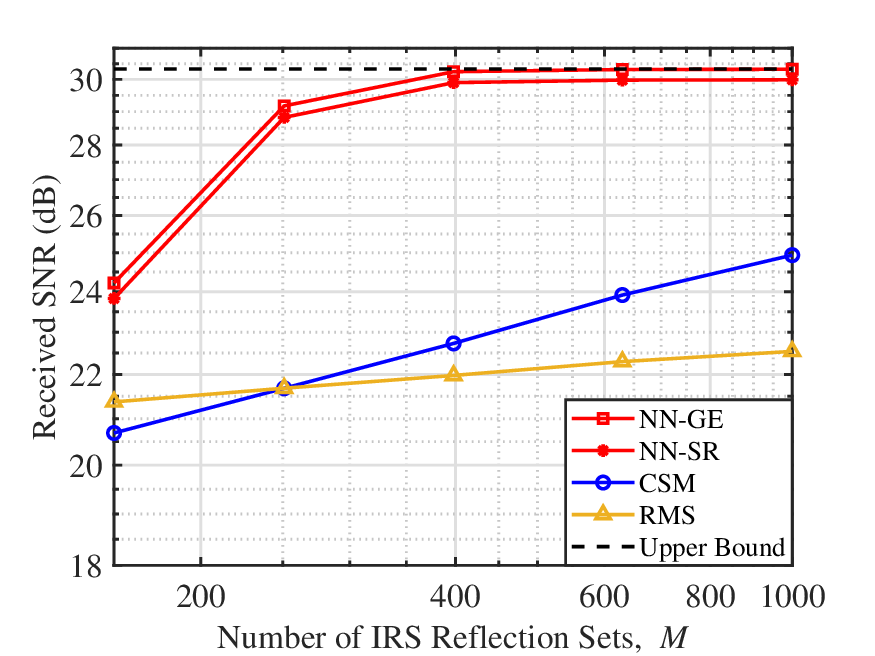} 
\caption{Received SNR versus the number of IRS reflection sets with $K=1$ and $\alpha=2$. }
\label{fig2}\end{center}
\vspace{-10pt}
\end{figure}
\subsection{Simulation Results}
We evaluate the received SNR by different schemes in both single-user and multiuser scenarios. For the proposed scheme, we first use the single-layer NN to estimate ${{{\boldsymbol{h}}}}_k, k \in {\cal{K}}$, and then apply the geometry-based method and the SDR method to optimize the IRS passive reflection in the single-user and multiuser scenarios, respectively (labeled as ``NN-GE'' and ``NN-SDR''), as presented in Section \ref{SECIIIC}. In addition, for both scenarios, we also show the performance by directly applying the successive refinement method, where the IRS passive reflection is initialized based on RMS given in (\ref{s3001}) (labeled as ``NN-SR'').

\begin{figure}[t]
\begin{center}
\centering
\includegraphics[scale=0.5]{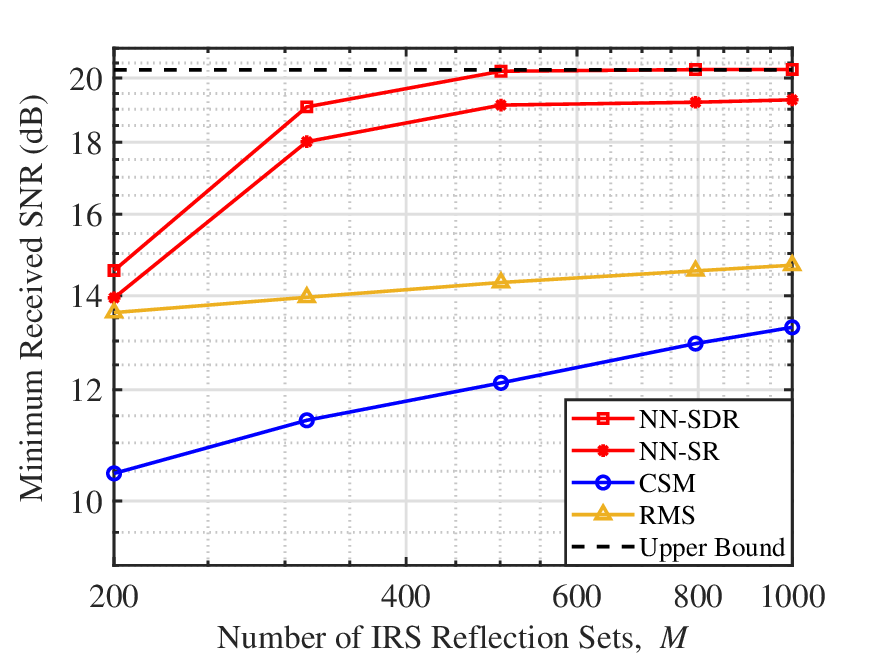} 
\caption{Minimum received SNR versus the number of IRS reflection sets with $K=5$ and $\alpha=1$. }
\label{fig3}\end{center}
\vspace{-10pt}
\end{figure}

\begin{figure}[t]
\begin{center}
\centering
\includegraphics[scale=0.5]{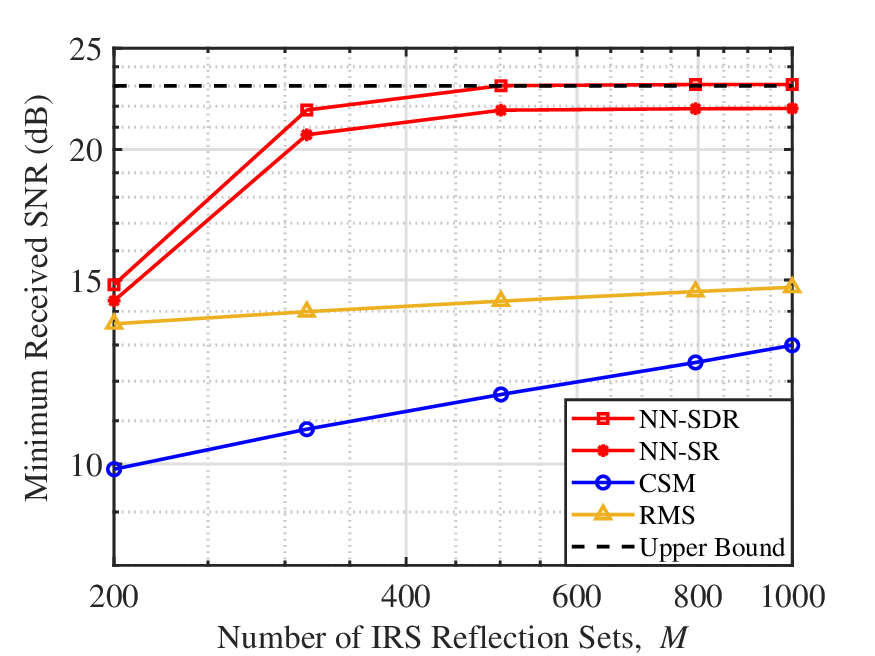} 
\caption{Minimum received SNR versus the number of IRS reflection sets with $K=5$ and $\alpha=2$.}
\label{fig4}\end{center}
\vspace{-10pt}
\end{figure}

First, Fig. \ref{fig1} and Fig. \ref{fig2} show the received SNR under different schemes in the single-user case with the number of controlling bits for IRS phase shifts $\alpha=1$ and $\alpha=2$, respectively. It is observed that both the NN-GE and NN-SR methods significantly outperform the benchmark schemes by fully exploiting the users' power measurements for channel estimation. In particular, with increasing $M$, the performance of our proposed scheme quickly converges to the upper bound achievable with perfect CSI. Moreover, the SNR performance improves by increasing $\alpha$ from 1 to 2, as expected, thanks to the higher phase-shift resolution for both channel estimation and reflection design. Furthermore, the small gap between the NN-SDR and NN-SR schemes demonstrates that the IRS passive reflection can be more efficiently optimized with linear complexity over $N$ if small performance loss is tolerable.
Next, Fig. \ref{fig3} and Fig. \ref{fig4} show the minimum received SNR among $K=5$ users with $\alpha=1$ and $\alpha=2$, respectively. Similar observations for the single-user case can be made for the multiuser case. In addition, it is observed that CSM performs worse than RMS in the multiuser case, due to its inefficacy to adapt to more complex utility functions such as that given in (\ref{s3002}) when $K>1$.

\section{Conclusion}
In this paper, we proposed a new IRS channel estimation method based on user received signal power measurements with randomly generated IRS reflections, by exploiting a simple single-layer NN formulation. Numerical results showed that the IRS passive reflection design based on the estimated IRS channels can significantly outperform the existing power measurement based schemes and approach the optimal performance assuming perfect CSI with significantly reduced power measurements in an IRS-aided multiuser communication system. The proposed IRS channel estimation and reflection optimization approach can be extended to other setups such as multi-antenna BS with adaptive precoding, multi-antenna user receivers, multiple IRSs, which will be studied in future work.

\bibliographystyle{IEEEtran}
\bibliography{mybib}

\end{document}